\documentclass[prl,aps,amssymb,amsmath,reprint,superscriptaddress,showpacs]{revtex4-1}
\usepackage{bm}
\usepackage{graphicx}
\usepackage{color}

\newcommand{\akd}{a^{\dagger}_{k}}
\newcommand{\ak}{a^{\phantom{\dagger}}_{k}}
\newcommand{\w}{\omega}

\newcommand{\beq}{\begin{equation}}
\newcommand{\eeq}{\end{equation}}
\newcommand{\beqa}{\begin{eqnarray}}
\newcommand{\eeqa}{\end{eqnarray}}

\newcommand{\ket}[1]{\left| #1 \right\rangle}
\newcommand{\bra}[1]{\left\langle #1 \right|}

\begin{document}

\title{
Stabilizing Spin Coherence Through Environmental Entanglement in 
\\ Strongly Dissipative Quantum Systems
}

\author{Soumya Bera}
\author{Serge Florens}
\affiliation{Institut N\'{e}el, CNRS and Universit\'e Grenoble Alpes, F-38042 Grenoble, France}
\author{Harold U. Baranger}
\affiliation{Department of Physics, Duke University, Durham, North Carolina 27708, USA}
\author{Nicolas Roch}
\affiliation{Laboratoire Pierre Aigrain, \'{E}cole Normale Sup\'{e}rieure,
CNRS (UMR 8551), Universit\'{e} Pierre et Marie Curie,
Universit\'{e} Denis Diderot, 24 rue Lhomond, 75231 Paris Cedex 05, France}
\affiliation{Institut N\'{e}el, CNRS and Universit\'e Grenoble Alpes, F-38042 Grenoble, France}
\author{Ahsan Nazir}
\affiliation{Blackett Laboratory, Imperial College London, London SW7 2AZ, United Kingdom}
\affiliation{Photon Science Institute, The University of Manchester, Oxford
Road, Manchester M13 9PL, United Kingdom}
\author{Alex W. Chin}
\affiliation{Theory of Condensed Matter Group, University of Cambridge,
J J Thomson Avenue, Cambridge, CB3 0HE, United Kingdom} 

\begin{abstract}
The key feature of a quantum spin coupled to a harmonic bath---a model
dissipative quantum system---is competition between oscillator potential energy
and spin tunneling rate. We show that these opposing tendencies cause
environmental entanglement through superpositions of adiabatic and
antiadiabatic oscillator states, which then stabilizes the spin coherence
against strong dissipation. This insight motivates a fast-converging variational 
coherent-state expansion for the many-body ground state of the spin-boson model, 
which we substantiate via numerical quantum tomography.
\end{abstract}

\date{\today}

\maketitle

The coupling of a quantum object to a macroscopic reservoir plays a fundamental role 
in understanding the complex transition from the quantum to the classical 
world.
The study of such dissipative quantum phenomena has deep implications 
across a broad range of topics in physics~\cite{Raimond}, quantum 
technology~\cite{nielsen}, chemistry~\cite{NitzanBook}, and
biology~\cite{LambertNP2013}. While quantum information
stored in the quantum subsystem alone is lost during the interaction with the
unobserved degrees of freedom in the reservoir, it is in principle preserved in
the entangled many-body state of the global system. The nature of this complete
wavefunction has received little attention, especially regarding the
entanglement generated among the reservoir states. 
However, ultrafast experiments on solid-state and molecular nanostructures,
including biological complexes, are increasingly able to probe the details of
environmental degrees of freedom, whose quantum properties - particularly in
non-perturbative regimes - may be key to understanding the device
characteristics~\cite{Plenio}. Our purpose here is to unveil
a simple emerging structure of the wavefunctions in open quantum systems, using
the complementary combination of numerical many-body quantum tomography and a
systematic coherent-state expansion that efficiently encodes the entanglement
structure of the bath. This combination of advanced techniques
reveals how non-classical properties of the macroscopic environment can stabilise 
quantum coherence with respect to a purely semiclassical response of the bath.

An archetype for exploring the quantum dissipation
problem~\cite{Leggett,Weiss,BreuerPetruccione} is to start with the simplest
quantum object, a two-level system describing a generic quantum bit embodied by
spin states $\left\{|\uparrow\rangle,|\downarrow\rangle\right\}$, 
and to couple it to an environment consisting of an infinite collection 
of quantum oscillators $\akd$ 
(with continuous quantum number $k$ and energy $\hbar \omega_k$). Quantum 
superposition of the two qubit states is achieved through a splitting 
$\Delta$ acting on the transverse spin component, while dissipation (energy exchange 
with the bosonic environment) and decoherence 
are provided by a longitudinal interaction term $g_k$ with each displacement
field in the bath. 
This leads to the Hamiltonian of the celebrated continuum spin-boson model (SBM)~\cite{Leggett,Weiss}:
\begin{equation}
H = \frac{\Delta}{2} \sigma_x - 
\sigma_z \sum_k \frac{g_k}{2} (\akd+\ak) + \sum_k \w_k \akd \ak,
\label{ham}
\end{equation}
where we set $\hbar=1$, and the sums can be considered as integrals by introducing the 
spectral function of the environment, $J(\omega)\equiv\sum_{k}g_{k}^{2}
\delta(\omega-\omega_{k})$.
The generality of the SBM makes it a key model for
studying non-equilibrium dynamics, non-Markovian quantum evolution, 
biological energy transport, and the preparation and control of exotic quantum states in 
a diverse array of 
systems~\cite{scholes2011lessons,Leggett,Weiss,BreuerPetruccione,NitzanBook}.

The possibility of maintaining robust spin superpositions in the ground and 
steady states of the SBM has attracted considerable attention, 
primarily due to its 
implications for quantum computing \cite{Jennings09,RaussendorfBriegel2001}.
Previous numerical approaches have focused on observables
related to the qubit degrees of freedom
\cite{NRG-RMP08,Makri95,WangThoss08,NalbachThorwart10,WinterBulla09,
AlvermannFehske09,PriorPlenio_EffSim10,Florens_DissipSpinDyn11}, 
whilst a description of the global system-environment wavefunction has been confined
to variational studies \cite{EmeryLuther,Silbey,Chin_SubohmSBM11,Nazir,Demler}. 
Variational theory readily predicts the formation of semiclassical polaron states, 
which involve the adiabatic response of the environmental modes to the spin
tunneling, and thus the generation of strong entanglement between the qubit and the bath.
However, we shall demonstrate here
that the ground state of Hamiltonian~(\ref{ham}) contains additional 
non-classical correlations among the environmental oscillator
modes arising from their \emph{non}-adiabatic response to spin-flip
processes. We find that this entanglement structure is key 
for the stabilization of qubit superpositions relative to the
semiclassical picture, and, in addition, follows naturally from a systematic variational 
framework beyond the adiabatic polaron approximation.

\begin{figure}[t!]
\includegraphics[width=1.0\linewidth]{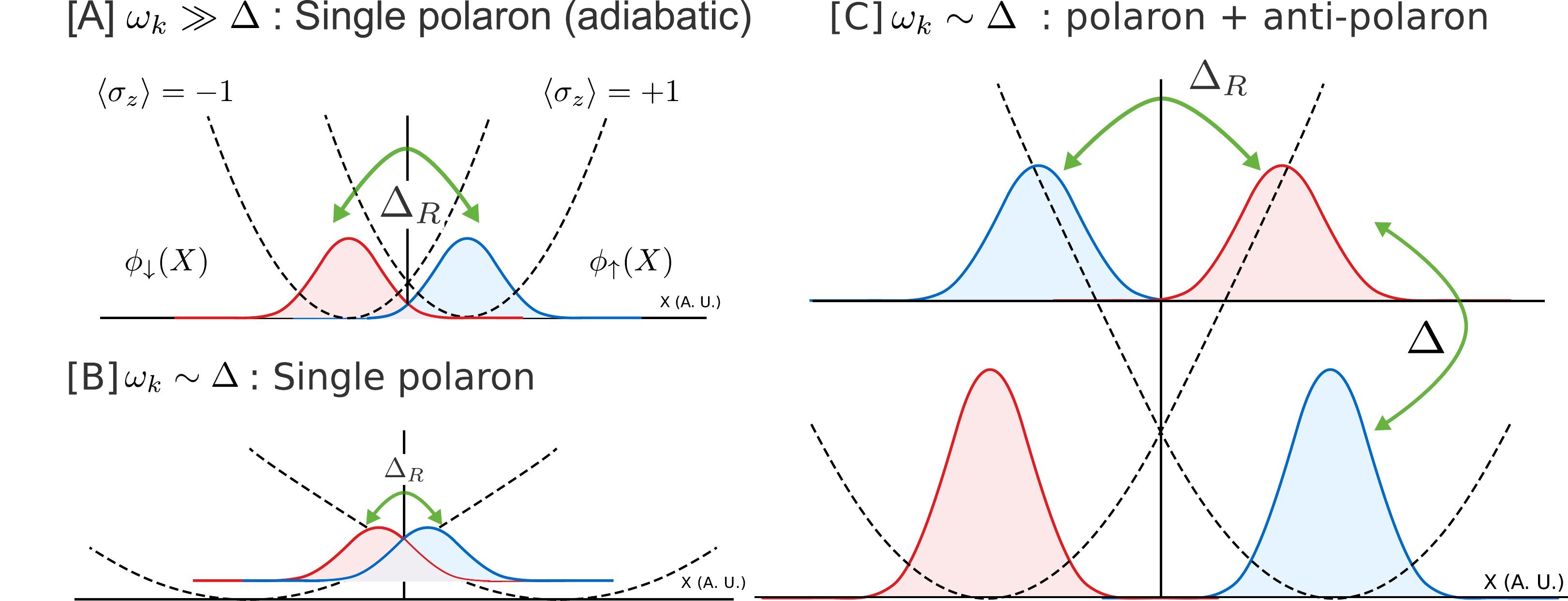}
\caption{(color online) 
Origins of polaron and antipolaron displacements in environmental wavefunctions.
Black dashed lines are the spin-dependent potential energies of a single
harmonic oscillator in the absence of spin tunneling ($\Delta=0$), while blue 
(red) curves sketch the single-mode wavefunctions of the oscillator (in real space $X$) on
the $\langle \sigma_z\rangle=1~(-1)$ potential surfaces. 
[A] High frequency modes ($\omega \gg\Delta$) tend to rest in the bottom of 
their (spin-dependent) potential due to the large energy of other displacements.
This leads to the formation of adiabatic polarons.
[B-C] Low frequency modes ($\omega \ll\Delta$) have shallow potentials with
well-separated minima. The inter-minima wavefunction overlap is reduced while
the potential cost of other displacements becomes less prohibitive. The
oscillators can either climb up the potential landscape, gaining some tunneling
energy (panel [B]), or become superposed with oppositely displaced
states---``antipolarons''---so that tunneling and potential energy are both
optimal (panel [C]).  Superposition of such polaron and antipolaron states
generates multimode entanglement in the complete environmental wavefunction.}
\label{antipolaron}
\end{figure}

In order to enlighten the nature of these emergent non-classical environmental states,
we start by analyzing the SBM with qualitative arguments based on energetics.
First, in the absence of tunneling, $\Delta=0$, the ground state of $H$ 
in Eq.~(\ref{ham}) is doubly degenerate. In the corresponding wavefunctions, 
the oscillators displace classically, in a direction that is fully correlated 
with the spin projection (adiabatic response):
$|\Psi_{\uparrow}\rangle=
|\!\uparrow\rangle\otimes |+f^{\mathrm{cl.}}\rangle$ and 
$|\Psi_{\downarrow}\rangle= |\!\downarrow\rangle\otimes |-f^{\mathrm{cl.}}\rangle$.
Here we introduce the product of semiclassical coherent states (displaced oscillators) 
$|\pm f\rangle \equiv e^{\pm\sum_k f_k (\akd-\ak)}|0\rangle$,
with the classical displacements $f_k^{\mathrm{cl.}}=\pm g_{k}/2\omega_{k}$ that shift 
each oscillator to the minimum of its static spin-dependent potential. 
This potential is evident in Eq.\,(\ref{ham}) for $\Delta=0$ and is shown explicitly 
in Fig.~\ref{antipolaron}A.

For $\Delta$$\neq$0, the oscillators experience a competition between spin 
tunneling and oscillator displacement energy. For high frequency modes ($\w_k\gg
\Delta$), transitions to other oscillator states 
are suppressed by the steep curvature of the potential (Fig.~\ref{antipolaron}A). 
Thus, these oscillators adiabatically tunnel with the spin between potential 
minima; the displacement of the oscillators reduces their overlap, suppressing 
the tunneling amplitude to a value $\Delta_R\ll\Delta$.
Extending this argument to low frequency modes ($\w_k \ll\Delta$) reveals a 
problem: the large separation of the minima ($g_k/2\w_k$) causes poor wavefunction overlap 
that prevents tunneling of the spin, thus destroying spin superposition. 
The classic scenario to overcome this problem is to adjust the displacements to 
smaller values, sacrificing potential energy to maintain spin-tunneling energy 
through better overlap (Fig.~\ref{antipolaron}B).  
Here, we argue for a new alternative scenario: because the potential surface is 
shallow for low frequency modes, transitions to other oscillator states become 
possible. Indeed, it is favorable for the oscillator wavefunction to include 
superpositions of coherent states with displacements \emph{opposite} to those 
dictated by the spin. In that way, direct tunneling transitions between the two 
potential surfaces are favored whilst keeping the main weight of the 
wavefunctions at low energy. We call these oppositely displaced oscillators 
``antipolaron'' states.

The strong competition between spin tunneling and oscillator displacement cannot
indeed be fulfilled by a single coherent state, even if optimized variationally.
The latter has been pursued in numerous variational
studies~\cite{EmeryLuther,Silbey,Chin_SubohmSBM11,Nazir,Demler},
embodied by the so-called Silbey-Harris (SH) Ansatz for the ground state of the
spin-boson model \cite{EmeryLuther,Silbey}:
\begin{equation}
\label{SH}
\big|\Psi^\mathrm{SH}\big> =
\big|\uparrow\big>
\otimes \big|\!+\!f^{\mathrm{SH}}\big> 
- \big|\downarrow\big> 
\otimes \big|\!-\!f^{\mathrm{SH}}\big>,
\end{equation}
where the displacements $f^{\mathrm{SH}}_k=[g_k/2]/(\w_k+\Delta_R)$ are determined by 
the variational principle. (Note that this Ansatz respects the symmetries of the Hamiltonian in the 
absence of a magnetic field along $\hat{z}$.) While this simple state possesses virtues, 
such as an accurate estimate of the renormalized tunneling frequency $\Delta_R=\Delta 
e^{-2\sum_k(f_k^\mathrm{SH})^2}$, it also has severe defects, such as spurious 
transitions~\cite{Nazir,mccutcheon11,lee12,chin06,chen08} and a drastic underestimation of 
the qubit coherence $\big< \sigma_x\big>$. Further works aiming at refining the 
variational Ansatz focused on the simple single-mode
case~\cite{stolze90,ren05,hwang10}, or were restricted to a non fully-optimal
variational state~\cite{zheng13}.
In light of the above discussion, the defects of the SH Ansatz are readily traced back to the 
lack of bath entanglement in wavefunction~(\ref{SH})---only the polaronic response 
is encoded in $f^{\mathrm{SH}}_k$. To capture the missing 
antipolaronic contributions and so the complete entanglement structure of the 
bath, we propose here a systematic coherent-state expansion of the many-body 
ground state:
\begin{equation}
\label{trial}
\big|\Psi\big> = \sum_{n=1}^{N} C_n \left[
\big|\uparrow\big> \otimes \big|\!+\!f^{(n)}\big> 
- \big|\downarrow\big> \otimes \big|\!-\!f^{(n)}\big> 
\right],
\end{equation}
with $ \big|\pm f^{(n)}\big>=e^{\pm\sum_k f^{(n)}_k (a_k^{\dagger}-a_k)}
\big|0\big>$ the $n^\mathrm{th}$ coherent state appearing in the wavefunction.
As is well known in many-body and chemical physics, variational methods may be greatly 
improved w.r.t. convergence by optimized basis choices. For the spin boson model, coherent 
states are naturally selected, as was understood from the $\Delta=0$ limit, and
are in addition very simple to parametrize in terms of displacements. 
As we will see, the coherent state expansion allows convergence to be achieved with
far fewer variational parameters than a brute-force variation of the
(exponentially large) coefficients of a full configuration (Fock) basis of
the environment states. This is due to the energetic constraints discussed
above, which strongly reduce the phase space volume of the allowed displacements
$f^{(n)}_{k}$. Indeed, for $\w_k>\Delta_R$, each displacement
function $f^{(n)}_{k}$ (for fixed $n$) will undergo quantum fluctuations between
the polaronic and antipolaronic branches, $f^{\mathrm{pol.}}_{k}\simeq
g_k/(2\w_k)$ and $f^{\mathrm{anti.}}_{k}\simeq -g_k/(2\w_k)$, respectively.
The main freedom in fixing a given displacement $f^{(n)}_k$ is then
in determining the crossover frequency from polaron to antipolaron behavior.
As a consequence (see below), physical properties are very precisely determined for 
moderate values of $N$, the number of coherent states involved in wavefunction~(\ref{trial}).

We can now readily understand how the emergence of environmental entanglement 
preserves spin coherence at strong dissipation. In the single-polaron Silbey-Harris
theory, the spin coherence $\big<\sigma_x\big> \simeq e^{- 2\sum_{k}(f^{\mathrm{SH}}_{k})^2} 
= \Delta_R/\Delta$ is incorrectly controlled by the exponentially small renormalized 
tunneling frequency $\Delta_R$. In contrast, the general wavefunction~(\ref{trial}) 
contains additional contributions to $\big<\sigma_x\big>$ of the type 
$e^{-\frac{1}{2}\sum_{k}(f^{(n)}_{k}+f^{(m)}_{k})^2}$ for $n\neq m$.
Quantum fluctuations that favor antipolarons will flip the sign of one
displacement with respect to the other, reducing the value of
the sum, and drastically increasing the exponential with
respect to the strongly suppressed value $\Delta_R/\Delta$. Thus, 
environmental correlations built into a multi-mode Schr\"odinger cat state affect 
the qubit properties in a dramatic way.
 
\begin{figure}[t!]
\includegraphics[width=0.85\columnwidth]{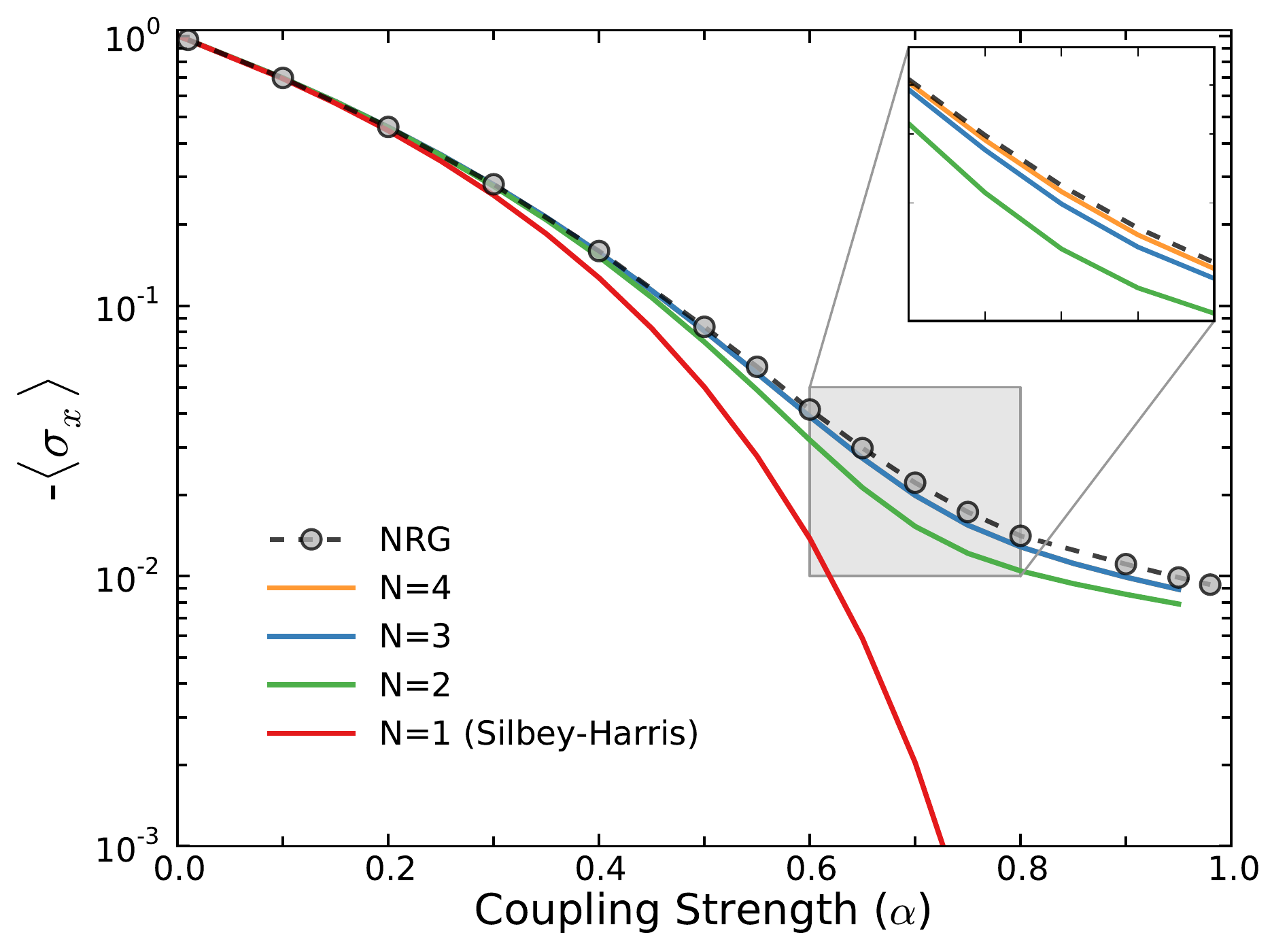}
\includegraphics[width=0.85\columnwidth]{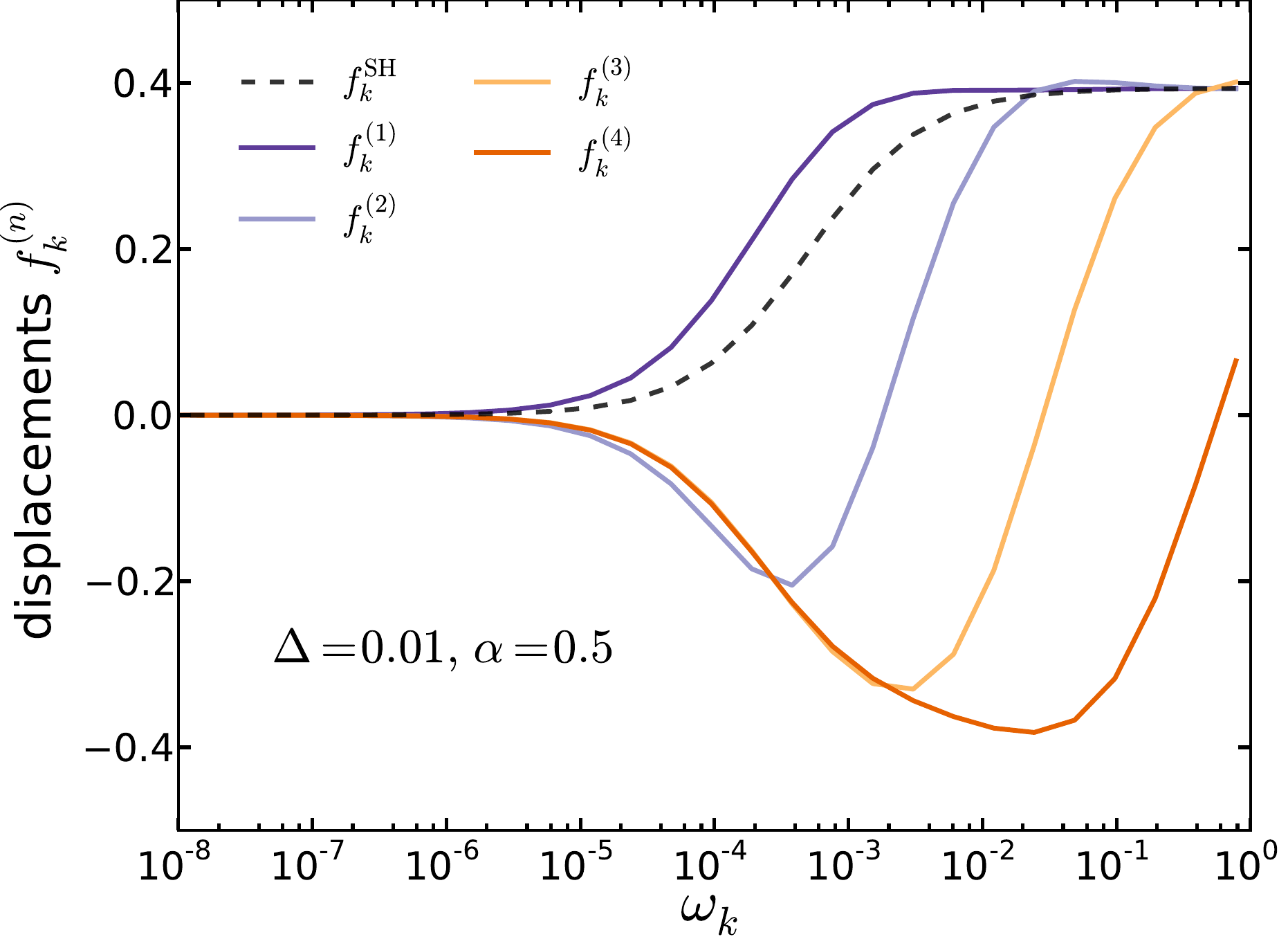}
\caption{Upper panel: Ground state coherence $-\big<\sigma_x\big>$ as a function
of dissipation strength $\alpha$ computed with the NRG (circles) for $\Delta/\w_c=0.01$ 
and compared to the results of the expansion Eq.~(\ref{trial}) with $N=1,2,3,4$
coherent states. The inclusion of an antipolaronic component to the wavefunction
($N\ge2$) has a drastic effect on the spin coherence. 
Lower panel: Displacements determined at order $N=1$ (dashed line) and $N=4$
(solid lines), showing the emergence of three antipolaron states at low energies, which
merge smoothly onto the polaron state at high energy (adiabatic regime)~\cite{Footnote1}.
Parameters are $\alpha=0.5$ and $\Delta/\w_c=0.01$.}
\label{ManyModesSx}
\end{figure}
We now turn to calculating the wavefunction Eq.~(\ref{trial}), 
where the displacements $f^{(n)}_{k}$ and coefficients $C_n$ are
determined by minimizing the total ground state energy 
$E=\big<\Psi| H| \Psi\big> / \big<\Psi| \Psi\big>$ for a fixed number
$N$ of coherent states ($1\leq n\leq N$)~\cite{supplement}. 
We focus here on the standard case of Ohmic dissipation \cite{Leggett,Weiss},
although our results should apply similarly to any type of spectral density.
The continuous bath of bosonic excitations then assumes a linear spectrum in frequency, 
$J(\w)\equiv\sum_{k}g_{k}^{2}\delta(\omega-\omega_{k}) = 2 \alpha \w \theta(\w_c-\w)$,
up to a high frequency 
cutoff $\w_c$, while the dissipation strength is given by
the dimensionless parameter $\alpha$. As a key check on the variational
solution, we carry out an exact non-perturbative solution of the SBM using the
numerical renormalization group (NRG)~\cite{Bulla}; for example, the spin
coherence $\big<\sigma_x\big>$, to which we shall compare the variational result,
is shown in Fig.~~\ref{ManyModesSx}.

The variational principle leads to a set of displacements $f^{(n)}_k$, shown in 
Fig.~~\ref{ManyModesSx}, that corroborate the physical picture given above: in 
addition to a fully positive displacement function $f^{(1)}_k$ (akin to the 
Silbey-Harris $f^{\mathrm{SH}}_k$ albeit quantitatively different), we find that 
all the other displacements undergo a crossover from positive value at high 
frequency to negative at low frequency. The total wavefunction~(\ref{trial}) is 
thus strongly entangled.

This environmental entanglement drastically affects the spin coherence
$\big<\sigma_x\big>$: note the difference in Fig.~\ref{ManyModesSx} between the
$N=2$ and $N=1$ solutions. In the latter, the coherence is given by the tiny
renormalized qubit frequency, $\Delta_R=\Delta (\Delta
e/\w_c)^{\alpha/(1-\alpha)}$ for $\Delta/\w_c\ll1$.  As the number of polarons
increases, however, the variational solution rapidly converges to the exact NRG result,
even capturing the saturation at strong dissipation
$-\big<\sigma_x\big>=\Delta/\w_c$ for $\alpha\to1$~\cite{LeHurReview}. These
panels thus confirm one important message of our study: emergent entanglement within
the environment stabilizes coherence of the spin, a result that is robust with
respect to coupling the whole system to a low temperature thermal bath~\cite{supplement}.

\begin{figure}[t]
\includegraphics[width=1.02\linewidth]{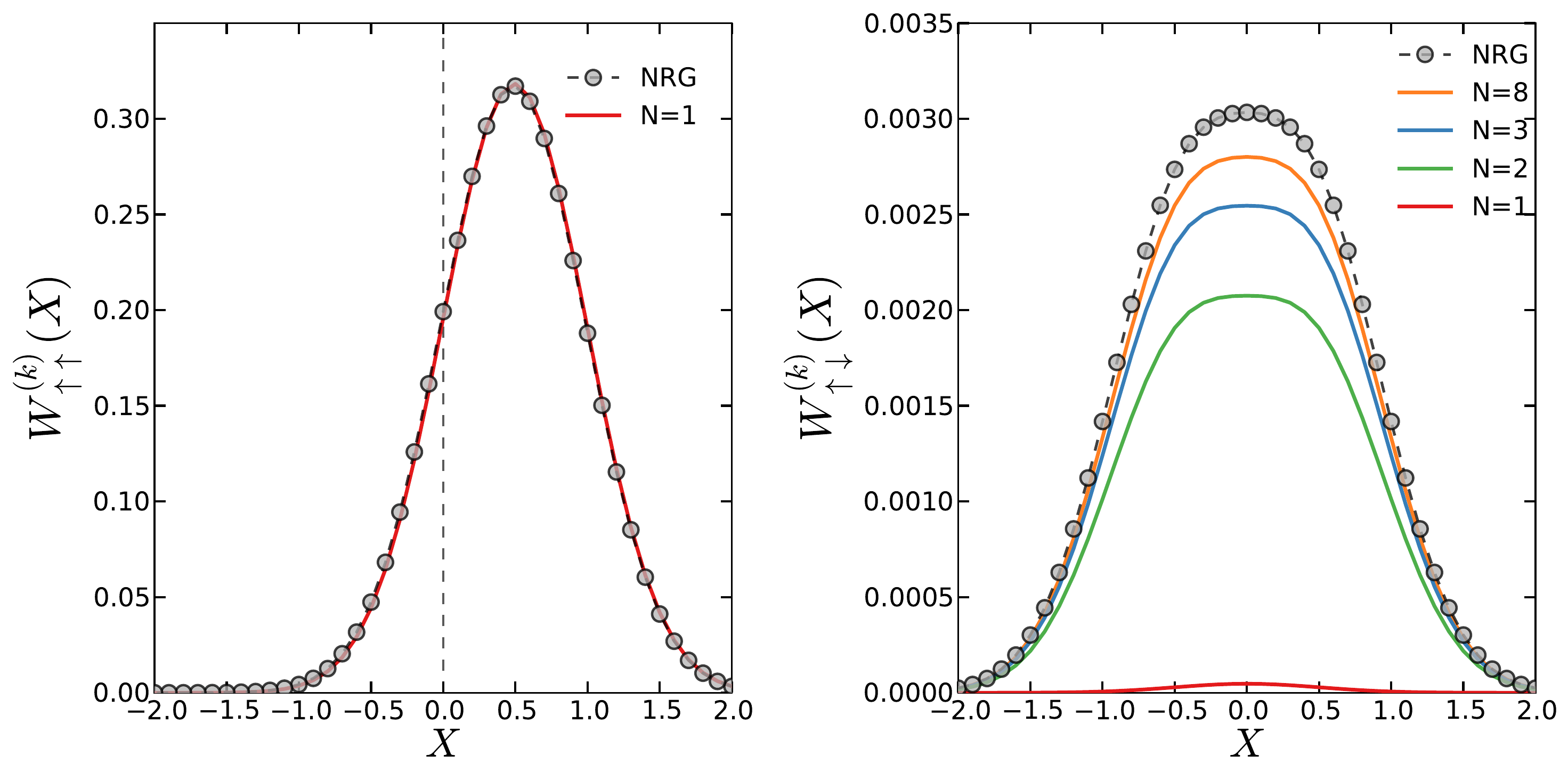}
\caption{Left: Spin-diagonal Wigner distribution, as obtained from the NRG,
computed for a high energy mode. Polaron formation leads to a classical
shift $X=f_k^{\mathrm{cl.}}$, that is fully captured by the Silbey-Harris
state (single coherent state). Right: The spin-off-diagonal Wigner function
is badly approximated by the single polaron $N=1$ state, as 
antipolaron contributions dominate in this quantity.
A single antipolaron ($N=2$) quickly restores the correct magnitude,
and reveals a pair of shifted Gaussians, as expected (see text).
[Parameters here are $\alpha=0.8$ and $\Delta/\w_c=0.01$.] 
}
\label{ManyModesWigner}
\end{figure}

We finally provide firm support for the above scenario by developing a 
many-body quantum tomography technique that allows direct characterization of the
ground-state wavefunction based on non-perturbative NRG computations. While one cannot
plot the complete many-body wavefunction, aspects can be accessed via standard
Wigner tomography~\cite{Raimond,ReviewTomography} a technique which has
witnessed impressive experimental developments lately in the field of
superconducting circuits \cite{Hofheinz,Eichler1,Eichler2}.  We choose
to trace out all modes except the qubit degree of freedom together with a single
bath mode with quantum number $k$.  Projecting first onto only the
$\big|\uparrow\big>$ part of the wavefunction, we obtain the Wigner function
$W_{\uparrow\uparrow}^{(k)}(X)$ as a function of the displacement $X$ of
oscillator $k$. For the wavefunction~(\ref{trial}), this has a straightforward
interpretation~\cite{Raimond,supplement}: 
the probability in phase space is
simply the sum of Gaussian peaks centered at $X\simeq (f^{(n)}_k+f^{(m)}_k)/2$.
For high energy modes (adiabatic regime), all displacements are
polaronic, $f^{(n)}_k\simeq f^{\mathrm{cl.}}_k = g_k/(2\w_k)$, so that a single 
shifted Gaussian appears in $W_{\uparrow\uparrow}^{(k)}(X)$, see  
Fig.~\ref{ManyModesWigner}(a). A single coherent state [{\it i.e.} the
Ansatz~(\ref{SH})] is sufficient in this high frequency regime to reproduce 
the NRG data perfectly, demonstrating the presence of polarons in the wavefunction. 

Antipolarons appear more clearly in the spin off-diagonal ground state Wigner function
$W_{\uparrow\downarrow}^{(k)}(X)$, rather than in $W_{\uparrow\uparrow}^{(k)}(X)$.
Note that such conditional Wigner tomography was considered for instance in recent
measurements of the moments $\big<(a^\dagger)^n (a)^m \sigma^i\big>$ for a
carefully prepared state entangling microwave photons and a superconducting
qubit in a circuit QED experiment~\cite{Eichler2}.
In $W_{\uparrow\uparrow}^{(k)}(X)$, the antipolarons are hidden because their weights 
$C_n$ tend to be smaller than that of the main polaron. In contrast,
$W_{\uparrow\downarrow}^{(k)}(X)$ is governed by cross polaron-antipolaron
contributions which peak at $X\simeq \pm (f^{(n)}_k-f^{(m)}_k)/2$~\cite{supplement}; 
other terms of
polaron-polaron type have an exponentially small weight of order $\Delta_R$.
The emergence of antipolaronic, namely opposite, displacements in $f^{(n)}$ and
$f^{(m)}$ should thus appear as a pair of symmetric Gaussians in
$W_{\uparrow\downarrow}^{(k)}(X)$. This is indeed observed in the NRG data for
intermediate frequencies, when adiabatic and antiadiabatic entanglement is
maximal, as shown in Fig.~\ref{ManyModesWigner}(b). In this case, a single
coherent state (fully polaronic) completely fails, but our
expansion~(\ref{trial}) quickly converges to the NRG results.

In conclusion, we have shown how environmental entanglement emerges in the ground state 
wavefunction of the spin-boson model, and why it surprisingly has a dramatic influence
on the qubit coherence. This understanding led us to develop a general framework 
to rationalize many-body wavefunctions in strongly interacting open quantum 
systems. Proposals have been made to realize the strongly dissipative 
spin-boson model in various physical systems~\cite{Recati,Porras}, most notably
by coupling a superconducting qubit to Josephson junction
arrays~\cite{LeHur,Goldstein}, giving hope that experimental studies of 
environmental states should become accessible in the near future.
The advances made in the present work opens the door to a better
understanding of several interesting issues, such as photon transport in
dissipative models~\cite{LeHur,Goldstein}, quantum phase transitions for sub-Ohmic 
baths~\cite{WinterBulla09,Bulla}, and studies of biased spin-boson
systems~\cite{Nazir}.

S.B., S.F., and H.U.B. thank the Grenoble Nanoscience Foundation for
funding under RTRA contract CORTRANO, and support by US DOE, Division of 
Materials Sciences and Engineering, under Grant No.\,DE-SC0005237.
A.N. thanks Imperial College and the University of Manchester for support. 
A.W.C. acknowledges support from the Winton Programme for the Physics of 
Sustainability.

\newpage
\setcounter{figure}{0}
\setcounter{table}{0}
\setcounter{equation}{0}

\onecolumngrid

\vspace{1.0cm}
\begin{center}
{\bf \large Supplementary information for
``Stabilizing spin coherence through environmental entanglement in strongly
dissipative quantum systems"}
\vspace{0.5cm}\\

\begin{quote}
{
\small
We present here basic technical details on the different methods used in the
main text. In the first part, we derive from the coherent state expansion 
various physical quantities of interest: ground state energy, spin coherence
and single-mode Wigner functions. In the second part, 
we provide information on how the (standard) numerical renormalization group simulations 
were set up in order to perform the Wigner tomography.
Finally, the third part presents finite temperature calculations, which indicate that 
the emergence of antipolarons is robust to thermal effects.
}
\end{quote}

\end{center}

\global\long\def\theequation{S\arabic{equation}}

\global\long\def\thefigure{S\arabic{figure}}

\section{General coherent state expansion in open quantum systems}

\subsection{Energetics of the trial state}

We consider the unbiased spin-boson model,~\cite{SLeggett,SWeiss} as defined by the Hamiltonian 
(1) of the main text: 
\begin{equation}\label{Hspinboson}
H=\frac{\Delta}{2}\sigma_x+\sum_k\omega_k
a_k^{\dagger}a_k-\sigma_z\sum_k\frac{g_k}{2}(a_k^{\dagger}+a_k), \end{equation}
with tunneling energy $\Delta$, a set of oscillator frequencies $\omega_k$, and
system-oscillator coupling strengths $g_k$ (assumed real). Here,
$\sigma_z=\ket{\uparrow}\bra{\uparrow}-\ket{\downarrow}\bra{\downarrow}$, with
spin basis states $\ket{\downarrow}$ and $\ket{\uparrow}$, and $a_k^{\dagger}$
($a_k$) is the oscillator creation (annihilation) operator for mode $k$.
Hamiltonian~(\ref{Hspinboson}) spans the cases of few discrete modes up to a continuum
of bosonic fields, in which case the discrete $k$-sum ought to be replaced
by an integral over energy.

The ground state wavefunction is expanded on an infinite set of
coherent states, as discussed in the main text: 
\begin{equation}
\label{Strial}
\big|\Psi\big> = \sum_{n=1}^{\infty} C_n \left[
\big|\uparrow\big> \otimes \big|\!+\!f^{(n)}\big> 
- \big|\downarrow\big> \otimes \big|\!-\!f^{(n)}\big> 
\right],
\end{equation}
where the bosonic part of the wavefunction involves coherent states of the form
\begin{equation}
\ket{\pm f^{(n)}}=e^{\pm\sum_k f^{(n)}_k (a_k^{\dagger}-a_k)}\ket{0},
\end{equation}
defined as products of displaced states, where $\ket{0}$ represents all oscillators 
being in the vacuum state. The presence of a $\mathbb{Z}_2$ symmetry, namely 
($|\uparrow\big>\to|\downarrow\big>$, $|\downarrow\big>\to|\uparrow\big>$, $a_k\to-a_k$), 
and the need for minimizing the spin tunneling energy enforces the chosen relative sign
between the up and down components of the ground state wavefunction in
Eq.~(\ref{Strial}).

Displacements $f_k^{(n)}$ and weights $C_n$ are taken as free parameters (both
are real numbers), that will be varied to minimise the total ground state energy 
$E=\big<\Psi\big|H\big|\Psi\big>/ \big<\Psi\big|\Psi\big>$, with
\begin{eqnarray}
\big<\Psi\big|H\big|\Psi\big> &=& 
- \Delta \sum_{n,m} C_n C_m e^{-\frac{1}{2}\sum_q [f^{(n)}_q+f^{(m)}_q]^2}
+ \sum_{n,m} C_n C_m \sum_k 2 \w_k f^{(n)}_k f^{(m)}_k e^{-\frac{1}{2}\sum_q [f^{(n)}_q-f^{(m)}_q]^2}\\
&&- \sum_{n,m} C_n C_m \sum_k g_k [f^{(n)}_k+f^{(m)}_k] e^{-\frac{1}{2}\sum_q [f^{(n)}_q-f^{(m)}_q]^2}
\end{eqnarray}
and $\big<\Psi\big|\Psi\big> = 2\sum_{n,m} C_n C_m e^{-\frac{1}{2}\sum_q
[f^{(n)}_q-f^{(m)}_q]^2}$. 

In the case of a single coherent state, we recover the usual Silbey-Harris
energy functional:~\cite{SSilbey}
\begin{equation}\label{eSHgs}
E^{\rm{(SH)}}=-\frac{\Delta}{2}e^{-2\sum_k(f_k^{\rm (SH)})^2}+\sum_k\omega_k
(f_k^{\rm (SH)})^2-\sum_k g_k f_k^{\rm (SH)},
\end{equation}
with the displacement $f_k^{\rm (SH)}= [g_k/2]/(\w_k+\Delta_R)$ readily
obtained from the variational principle, $\partial E^{\rm{(SH)}} / \partial
f_k^{\rm (SH)} = 0$. 

\subsection{Ground state spin coherence}
The coherent-state expansion~(\ref{Strial}) for the wavefunction allows us to
compute in a simple way all physical observables. We start with the spin
coherence in the ground state, a quantity that was discussed in detail in
the main text. We readily find:
\begin{equation}
\big<\sigma_x \big> = - \frac{ 
\sum_{n,m} C_n C_m e^{-\frac{1}{2}\sum_q [f^{(n)}_q+f^{(m)}_q]^2}}
{ 2\sum_{n,m} C_n C_m e^{-\frac{1}{2}\sum_q [f^{(n)}_q-f^{(m)}_q]^2}}.
\end{equation}
This expression was used to determine the spin coherence in Fig.~2A of
the main text.

\subsection{Spin-diagonal single-mode Wigner function}

We discuss here the spin-dependent Wigner distributions associated to 
the reduced density matrix defined in the subspace spanned by the qubit and one 
{\it given} oscillator mode with frequency $\omega_k$. The qubit degrees of freedom
can be used for filtering out the polaron and antipolaron contributions within
the wavefunction, thanks to appropriate insertions of Pauli matrices in
the standard definition of the Wigner function.~\cite{SRaimond} We consider here
the projection onto the $\ket{\uparrow}$ component only:
\begin{equation}
\label{WDiagDef}
W^{(k)}_{\uparrow\uparrow}(X)=
 \frac{1}{\big<\Psi\big|\Psi\big>}
\int \!\!\! \frac{\mathrm{d^2}\lambda}{\pi^2}\;
e^{X(\bar{\lambda}-\lambda)} 
\big<\Psi\big| e^{\lambda \akd-\bar{\lambda}\ak}
\frac{1+\sigma_z}{2}
\big|\Psi\big>.
\end{equation}
This Wigner distribution can be evaluated using the coherent state
expansion of the ground state~(\ref{Strial}).
\begin{equation}
W^{(k)}_{\uparrow\uparrow}(X)=
 \frac{1}{\big<\Psi\big|\Psi\big>} \int \!\!\! \frac{\mathrm{d^2}\lambda}{\pi^2}\;
e^{X(\bar{\lambda}-\lambda)} 
\sum_{n,m} C_n C_m
\big<f^{(n)}\big| e^{\lambda \akd-\bar{\lambda}\ak}
\big|f^{(m)}\big>.
\label{SupWSzIntegral}
\end{equation} 
Using standard coherent state algebra, we easily obtain the required overlaps,
\begin{eqnarray}
\big<f^{(n)}\big| e^{\lambda \akd-\bar{\lambda}\ak}
\big|f^{(n)}\big> &=& 
e^{(\lambda-\bar\lambda) f^{(n)}_k}
e^{-\lambda\bar\lambda/2}\\
\big<f^{(n)}\big| e^{\lambda \akd-\bar{\lambda}\ak}
\big|f^{(m)}\big> &=& e^{-\frac{1}{2}\sum_q
(f_q^{(n)}-f_g^{(m)})^2}
e^{\lambda f^{(n)}_k- \bar\lambda f^{(m)}_k} 
e^{-\lambda\bar\lambda/2}\\
\big<f^{(m)}\big| e^{\lambda \akd-\bar{\lambda}\ak}
\big|f^{(m)}\big> &=& 
e^{(\lambda-\bar\lambda) f^{(m)}_k}
e^{-\lambda\bar\lambda/2}.
\end{eqnarray}
Performing the Gaussian integral in Eq.~(\ref{SupWSzIntegral}) yields:
\begin{equation}
W^{(k)}_{\uparrow\uparrow}(X)=
\frac{1}{\pi\big<\Psi\big|\Psi\big>}
\sum_{n,m} C_n C_m 
e^{-\frac{1}{2}\sum_{q\neq k} (f^{(n)}_{q}-f^{(m)}_{q})^2}
 e^{-2\big(X-\frac{f^{(n)}_{k}+f^{(m)}_{k}}{2}\big)^2}.
\label{SupWSzFinal}
\end{equation}
In the adiabatic limit $\w_k\gg\Delta$, all displacements become
classical, $f_k^{(n)}\simeq f_k^{\mathrm{cl.}} = g_k/(2\w_k)$, so that the Wigner
function reduces to a single Gaussian centered on $X\simeq g_k/(2\w_k)$,
as demonstrated in Fig.~3A of the main text.

\subsection{Spin-off-diagonal single-mode Wigner function}

In order to highlight the emergence of antipolaronic contributions in the
wavefunction, we now insert the off-diagonal $\sigma_x$ Pauli matrix in
the usual definition of the Wigner function:
\begin{equation}
\label{WOffDiagDef}
W^{(k)}_{\uparrow\downarrow}(X)=
 \frac{1}{\big<\Psi\big|\Psi\big>}
\int \!\!\! \frac{\mathrm{d^2}\lambda}{\pi^2}\;
e^{X(\bar{\lambda}-\lambda)} 
\big<\Psi\big| e^{\lambda \akd-\bar{\lambda}\ak}
\sigma_x
\big|\Psi\big>.
\end{equation}
From the trial state~(\ref{Strial}), we get:
\begin{eqnarray}
W^{(k)}_{\uparrow\downarrow}(X)=
 \frac{1}{\big<\Psi\big|\Psi\big>} \int \!\!\! \frac{\mathrm{d^2}\lambda}{\pi^2}\;
e^{X(\bar{\lambda}-\lambda)} 
\sum_{n,m} C_n C_m
\big<-f^{(n)}\big| e^{\lambda \akd-\bar{\lambda}\ak}
\big|f^{(m)}\big>.
\label{SupWSplus}
\end{eqnarray}
A computation similar to the one performed above leads to the final result:
\begin{equation}
\label{SupWSplusFinal}
W^{(k)}_{\uparrow\downarrow}(X)=
\frac{1}{\pi\big<\Psi\big|\Psi\big>}
\sum_{n,m} C_n C_m 
e^{-\frac{1}{2}\sum_{q\neq k} (f^{(n)}_{q}+f^{(m)}_{q})^2}
\left[
 e^{-2\big(X-\frac{f^{(n)}_{k}-f^{(m)}_{k}}{2}\big)^2}
 +e^{-2\big(X+\frac{f^{(n)}_{k}-f^{(m)}_{k}}{2}\big)^2}
\right].
\end{equation}
The above expression shows important differences from the $\ket{\uparrow}$-projected Wigner
distribution of Eq.~(\ref{SupWSzFinal}). First, Gaussian peaks form at position
near the difference of two displacements,
$X\simeq\frac{f^{(n)}_{k}-f^{(m)}_{k}}{2}$, so that the formation of antipolarons
results in finite displacements. Second, the main polaron contribution, $n=m=1$,
is suppressed by the tiny factor $e^{-2\sum_{q\neq k}
(f^{(1)}_{q})^2}\simeq\Delta_R/\Delta\ll 1$, in contrast to the spin-diagonal Wigner
function~(\ref{SupWSzFinal}), where it appears with a prefactor of order one.
Antipolarons are thus best resolved in the spin-off-diagonal Wigner function~(\ref{SupWSplusFinal}), 
as checked in Fig.~3B of the main text.

\section{Implementation of the NRG calculations}

The numerical solution of the spin-boson Hamiltonian~(\ref{Hspinboson})
with continuous spectrum relies on a logarithmic shell blocking of the 
bosonic modes onto energy intervals $[\Lambda^{-n-1}\w_c,\Lambda^{-n}\w_c]$ 
(with $\Lambda=2$ in all our calculations). This defines Wilson-shell
bosonic creation operators:
\begin{equation}
a^\dagger_n = \int_{\Lambda^{-n-1}\w_c}^{\Lambda^{-n}\w_c} \!\!\!\!\!\!\! 
\mathrm{d}k \; \akd \;.
\end{equation}
The resulting discrete Hamiltonian, which spans from arbitrarily small energy up
to the high energy cutoff $\omega_c$, is then iteratively diagonalised according
to the Numerical Renormalization Group (NRG) algorithm~\cite{SNRG-RMP08,SBulla}. 
One novel part of the NRG simulations performed for this work lies in the computation 
of the Wigner distribution reduced to the joint spin and single $k$-mode subspace.
In order to implement Eqs.~(\ref{WDiagDef}) and (\ref{WOffDiagDef}), we first
define arbitrary moments of the chosen oscillator with frequency 
$\w_k=\w_c\Lambda^{-n}$:
\begin{equation}
A_{\sigma_i;m,m'}^{(k)}=\big<\Psi| \sigma_i [a^\dagger_n]^m 
[a^{\phantom{\dagger}}_n]^{m'}|\Psi\big>,
\end{equation}
with $i=0,x,y,z$ labelling the Pauli matrices related to the spin projection (we
take $\sigma_0\equiv1$) and $m,m'$ positive integers.
Such ground state observables are readily computed within the NRG algorithm (for
typically
$0\leq m,m'<10$).
One can then expand Eqs.~(\ref{WDiagDef}) and (\ref{WOffDiagDef}) in a power series
in $\lambda$ and $\bar\lambda$, yielding
\begin{equation}
W_{\sigma_i}^{(k)}(X) = \frac{2}{\pi} \sum_{m,m'=0}^{+\infty}
\!\!\! A_{\sigma_i;m,m'}^{(k)} \frac{(-1)^{m+m'}}{m!m'!} 
\frac{\partial^{m+m'}}{\partial X^{m+m'}} e^{-2X^2} .
\end{equation}
The wanted Wigner distributions are now solely expressed in terms of the
NRG-computable moments $A_{\sigma_i;m,m'}^{(k)}$, and are shown in Fig.~3
of the main text.

\section{Finite temperature effects}
\begin{figure}[b]
\includegraphics[width=0.5\linewidth]{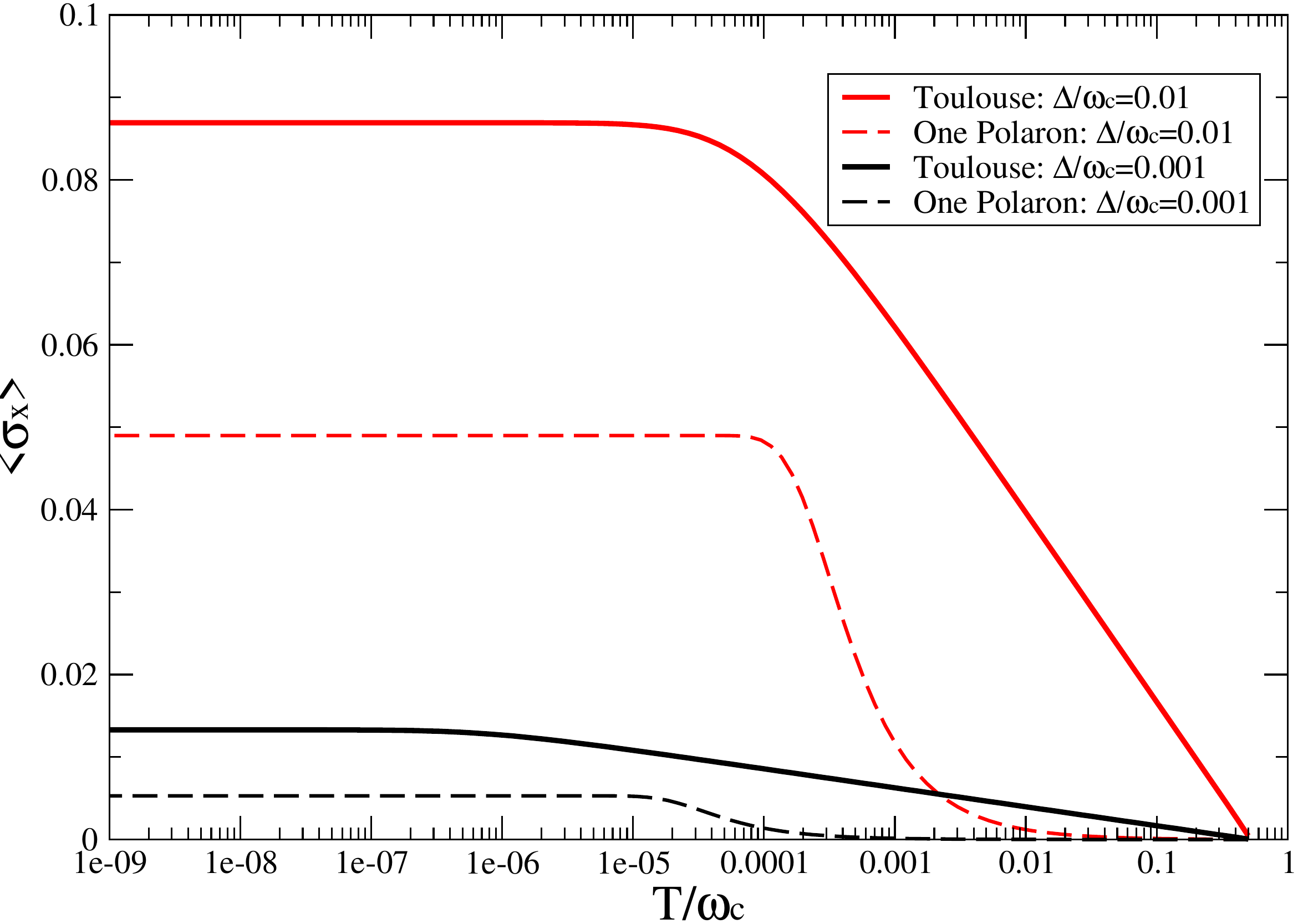}
\caption{Spin coherence at intermediate dissipation $\alpha=0.5$ as a function 
of temperature for $\Delta/\omega_c=0.001$ (black curves) and $\Delta/\omega_c=0.01$ 
(red curves), computed from the exact Toulouse solution (solid lines) 
and from the one-polaron Silbey-Harris theory (dashed lines).}
\label{thermal}
\end{figure}

One striking feature of the many-body ground state of the
spin-boson model is that the exact zero-temperature spin coherence $\big<\sigma_x\big>$ is
controlled at large dissipation by the bare scale 
$\Delta/\omega_c$, which is much larger than the renormalized
dimensionless tunneling energy $\Delta_R/\omega_c$ predicted by single-polaron
theory. This effect was shown in the main manuscript to arise from entanglement within the bath 
in the form of antipolaron contributions to the many body state.

We consider here the effect of finite temperature on the coherence, and
demonstrate that, within a single-polaron approach, the coherence is strongly
suppressed when the temperature reaches the scale $\Delta_R$, while it is preserved 
up to much higher temperatures (of the order of the bare $\Delta$) in the exact 
solution of the spin-boson model.
This indicates that antipolaron-type entanglement persists within the low-lying 
thermal excitations. We suppose here that an external thermal bath couples to our 
spin-boson system such that the system thermalizes with respect to many-body 
states of the total spin-boson Hamiltonian.

Generalizing the many-polaron Ansatz to capture the 
free energy of the spin-boson model is a formidable task beyond the scope 
of the present analysis.
In addition, we would like to have quantitative results for
$\big<\sigma_x\big>$ at finite temperature that are independent of the coherent
state expansion. We thus consider the solution of the spin-boson model
at the exactly-solvable Toulouse line, namely for the dissipation strength $\alpha=0.5$.
In this case, the spin-boson model can be mapped onto a model of non-interacting
electrons with a resonant level \cite{SLeggett,SWeiss}.  The free energy can then 
be computed from the standard expression:
\begin{equation}
F = - T \int_{-D}^{D} \!\!\! d\epsilon \; \log\left[1+e^{-\epsilon/T}\right] \rho(\epsilon),
\end{equation}
with $\rho(\epsilon)= (T_K/\pi)/[\epsilon^2+T_K^2]$ the effective electronic density of states 
of the equivalent resonant level model. Here, $D=4\omega_c/\pi$
denotes the corresponding fermionic bandwidth and $T_K=\Delta^2/\omega_c$
is the Kondo scale.
The spin coherence is readily obtained from the derivative of this free energy with
respect to the tunneling amplitude:
\begin{equation}
\big<\sigma_x\big> = - 2 \frac{d F}{d\Delta} = \frac{4\Delta}{\pi D} T 
\int_{-D}^{D} \!\!\! d\epsilon \;\log\left[1+e^{-\epsilon/T}\right]
\frac{\epsilon^2-T_K^2}{(\epsilon^2+T_K^2)^2}.
\end{equation}
This expression is numerically evaluated and shown as solid lines 
in Fig.~\ref{thermal} for two values of $\Delta$.

We now compare these exact results to the one-polaron theory, which can be extended 
to finite temperature,~\cite{SNazir} and provides a simple formula for the spin
coherence:
\begin{equation}
\big<\sigma_x\big>_{\mathrm{1pol.}} =\frac{\Delta_R}{\Delta}
\tanh\left(\frac{\Delta_R}{2T}\right).
\end{equation}
Thus, the coherence is rapidly suppressed by thermal effects in
the case of the single-polaron approximation when reaching a temperature 
that is of the order of the small scale $\Delta_R\propto T_K$,
see the dashed curves in Fig.~\ref{thermal}. Note in particular
how the exact solutions develop long tails at high temperature, rather than 
the sharp decay found in the single-polaron theory, a hallmark of many-body 
Kondo-type problems and antipolaron contributions in the present spin-boson context.
This comparison proves the relevance of antipolaron physics in the thermal
excitations of the spin-boson model.


\begin{thebibliography}{99}
\bibitem{Footnote1} 
In the NRG logarithmic discretisation of the bath spectrum, frequency points are 
unevenly spaced, leading to a saturation of $f_k$ at high frequencies, 
instead of the fall off obtained for a linear energy mesh.
\end{thebibliography}
\end{document}